\documentclass{article}

\PassOptionsToPackage{numbers, sort&compress}{natbib}
\usepackage[preprint]{neurips_2026}

\usepackage[utf8]{inputenc}
\usepackage[T1]{fontenc}
\usepackage{hyperref}
\usepackage{url}
\usepackage{booktabs}
\usepackage{amsfonts}
\usepackage{nicefrac}
\usepackage{microtype}
\usepackage{xcolor}

\usepackage{amsmath}
\usepackage{graphicx}
\usepackage{multirow}
\usepackage{array}

\usepackage{pifont}

\newcommand{\cmark}{\ding{51}}

\graphicspath{{../figs/}}

\title{\textbf{EduPanel: A Three-Agent LLM Judge for Teaching Videos --- Reliability, Complementarity, and Human Trust Calibration}}
\author{
Jia-Kai Dong$^{1}$ \And
Yi-Cheng Lin$^{1}$ \And
Hung-yi Lee$^{1,2}$ \\
\\
$^{1}$ National Taiwan University\\
$^{2}$ NTU Artificial Intelligence Center of Research Excellence \\
\texttt{b11901067@ntu.edu.tw}}

\begin{document}

\maketitle

\begin{abstract}
Teaching videos are becoming a major medium for education, creating a growing need for scalable evaluation of their pedagogical quality.
Existing automatic judges do not fully address this setting because teaching quality depends on multimodal evidence and should be evaluated with respect to the intended learner rather than as a universal property.
We present \textbf{EduPanel}, a rubric-grounded, learner-conditioned LLM judge that decomposes evaluation across specialized agents to produce interpretable assessments for different aspects of teaching quality.
Across expert studies, architecture ablations, and learner-persona analyses, EduPanel achieves reliability comparable to a median human expert. In expert evaluation, its feedback improves scoring accuracy (MAE 0.87→0.73), while experts remain able to detect unreliable outputs (AUC = 0.77) instead of accepting them blindly.
These results suggest that EduPanel can serve as an effective assistant for educational evaluation rather than a replacement for human experts.
\end{abstract}

\section{Introduction}

Instructional videos have become a major medium for education, from online courses and flipped classrooms to AI-generated educational content
\citep{sun2022preavatar,holmberg2025narrated,chen2025code2video,zhu2025paper2video}.
As AI lowers the cost of producing instructional videos, the supply of educational content is rapidly expanding.
This abundance creates a new challenge: learners and educators must increasingly identify which videos are pedagogically suitable for a particular learner, rather than merely find content that is factually correct.
Reliable automatic evaluation could provide scalable feedback, help identify videos requiring revision before deployment, and reduce the cost of expert review.
Human evaluation against pedagogical rubrics remains the gold standard
\citep{pianta2008class},
but it is difficult to scale.
Applying a pedagogical rubric requires trained raters, repeated viewing, and careful comparison between narration, visual presentation, learning objectives, and intended audience.
Moreover, many aspects of teaching quality are inherently subjective, limiting agreement even among experienced evaluators
\citep{pianta2008class}.
Large language models (LLMs) have therefore become increasingly popular as automatic judges of open-ended outputs \citep{zheng2023judging,liu2023geval,kim2024prometheus,gu2024survey}. 
Yet pedagogical evaluation differs from the text-quality settings where LLM judges have primarily been studied in two important ways. 
First, the signal is inherently \emph{multimodal}. 
Judgments about pacing, on-screen worked examples, and whether narration aligns with the visual presentation depend on what is \emph{shown}, not only what is said. 
Although recent video--language models can describe and reason over video content
\citep{nguyen2024video,xiong2024llavacritic}, existing LLM judges are not designed to incorporate such multimodal evidence.
Second, and central to this paper, pedagogical quality is inherently \emph{learner-dependent}: the same instructional material may be effective for one student but inappropriate for another. For example, a mathematically detailed explanation of neural networks may benefit advanced students but overwhelm beginners who lack the necessary prerequisites.
Effective teaching therefore depends on the learner's prior knowledge and expertise
\citep{Kalyuga01012003}, which a learner-independent judge cannot capture.

We introduce \textbf{EduPanel}, a multimodal, rubric-grounded judge that evaluates teaching videos with respect to a specified learner, rather than assigning a single learner-independent quality score.
Recent work has explored simulated learners for educational applications \citep{marquez2026simulating}; in contrast, we use learner profiles as an evaluation condition, allowing the same video to be assessed differently for different intended audiences.

Evaluating a teaching video requires several distinct forms of reasoning.
An evaluator must reconstruct and fact-check what the video teaches, assess its alignment with the course specification, and determine whether the instruction is appropriate for the target learner.
Rather than asking a single model to perform all of these tasks, EduPanel decomposes the evaluation across three specialized agents.
This decomposition follows the structure of the evaluation itself, allowing each agent to focus on a different source of evidence and type of judgment.
It also makes the evaluation more transparent: each score is accompanied by an inspectable content map, identified issues, and supporting rationale instead of a single opaque score.
Section~\ref{sec:judge} describes the architecture in detail.

Having built such a system, we argue that evaluating it should go beyond whether it
\emph{agrees} with human raters.
A practical judge should not only produce accurate scores, but also
\emph{complement} experts by identifying problems they miss rather than merely reproducing their judgments
\citep{wang2024unfair,dubois2024length,panickssery2024self,gu2024survey}.
Moreover, if the judge is intended to assist rather than replace experts, its commentary should serve as a reference signal that experts can critically verify, adopting useful suggestions while remaining able to detect incorrect ones.

We therefore evaluate EduPanel from three perspectives: evaluation reliability, design factors, and usefulness in expert workflows.

\paragraph{Research questions.}
\textbf{RQ1} (evaluation quality): How reliably does EduPanel evaluate teaching videos, and how does its behavior vary across rubric dimensions and information modalities?
\textbf{RQ2} (design validation): Which design choices, including video access, learner conditioning, and agent decomposition, contribute to EduPanel's behavior?
\textbf{RQ3} (practical utility): Does EduPanel provide useful and critically verifiable assistance in an expert evaluation workflow, including complementary error detection?

We answer these questions through the following contributions.

\paragraph{Contributions.}
\begin{enumerate}
\item We introduce \textbf{EduPanel}, an open-source, multimodal, rubric-grounded, learner-conditioned three-agent judge for teaching videos\footnote{\url{https://github.com/snooow1029/edupanel}}
(Sections~\ref{sec:judge},~\ref{sec:persona}).
\item  We show that learner conditioning and agent decomposition enable evaluation tailored to intended learners while providing inspectable intermediate evidence.
\item We provide a system-level evaluation of EduPanel, including per-dimension reliability analysis, persona sensitivity, and architecture and input ablations
(Sections~\ref{sec:rq1}--\ref{sec:persona}).

\item We evaluate EduPanel in a realistic expert workflow and show that its commentary improves expert scoring while contributing complementary observations. A planted-error study further shows that experts can distinguish unreliable outputs from reliable ones ($\mathrm{AUC}=0.77$), indicating that they use the judge critically rather than accepting it wholesale (Sections~\ref{sec:assist},~\ref{sec:safety}).
\end{enumerate}

Our experiments are conducted on a deliberately small, curated benchmark consisting of 12 videos, 10 evaluation dimensions, four subject domains, and one deployed backbone.

Accordingly, our conclusions concern the proposed system and its calibration rather than population-level estimates.

\section{Related Work}
\paragraph{From LLM judges to multimodal, multi-agent judges.}
LLMs are now widely used as judges of open-ended outputs, reaching high agreement with human preferences \citep{zheng2023judging,liu2023geval,kim2024prometheus}, though with known biases and typically meta-evaluated against curated, text-only benchmarks \citep{gu2024survey,lambert2024rewardbench}. 
Two extensions bring this closer to our setting. 

First, multimodal models enable judging beyond text-only outputs by incorporating visual and audio-visual signals
\citep{chen2024mllmjudge,xiong2024llavacritic}.
However, existing judges primarily target images or short clips, and to our knowledge no prior judge evaluates full-length teaching videos along \emph{pedagogical} dimensions.
Second, decomposing evaluation across \emph{role-differentiated agents} has been used to structure evaluative reasoning \citep{du2024debate,chan2024chateval}. 
We build on both threads but change both the evaluation \emph{target} and the purpose of the decomposition. 
Whereas prior judges estimate response quality as an absolute property, EduPanel performs learner-conditioned evaluation of educational videos by estimating \emph{pedagogical fitness}, namely quality relative to a specific learner. 
It uses role decomposition to improve \emph{auditability and role-level controllability} rather than aggregate accuracy.
It is studied not only as an evaluator but as a member of a human rating workflow.

\paragraph{Generating and assessing teaching content.}
A growing line of work focuses on automatically generating instructional and presentation videos from lecture material, slides, scripts, or research papers
\citep{sun2022preavatar,holmberg2025narrated,chen2025code2video,zhu2025paper2video}.
Several of these pipelines even introduce their own generation-time quality metrics, such as VLM-as-a-judge aesthetic scores and knowledge-recovery quizzes \citep{chen2025code2video,zhu2025paper2video}---but these ask whether information is \emph{conveyed}, not whether a video teaches well. 
Educational assessment proper, by contrast, has moved from human rubric instruments such as CLASS \citep{pianta2008class} to LLM-based rubric grading of student and teaching artifacts \citep{wang2023chatgpt,demszky2024feedback}, yet remains almost entirely \emph{text}-based (essays, transcripts, short answers) and graded against an \emph{absolute} rubric.
EduPanel sits in the gap these two lines leave open and differs on both axes: it scores the \emph{video} on dimensions that turn on what is shown, and it scores \emph{pedagogical fitness} for a specified learner rather than absolute quality---so it could in principle serve as an automatic quality gate or formative-feedback signal for the generation pipelines above, an application we do not evaluate here.

\paragraph{Personalized, learner-conditioned evaluation.}
Most automated assessment approaches evaluate instructional quality without explicitly modeling learner characteristics.
However, pedagogical research has long shown that instructional effectiveness depends on the learner's prior knowledge and expertise \citep{Kalyuga01012003}.
Recent work shows LLMs can role-play learners with specified backgrounds convincingly enough to stand in for them \citep{park2023generative}, a capability used in education to train teachers against synthetic students \citep{markel2023gpteach}. 
Closest to us, \citet{lu2024generative} simulate diverse student profiles to \emph{evaluate} the quality of static question items. 
We extend this evaluative use from static items to teaching videos: a persona-simulator agent scores each video relative to a target learner, so the judge
measures \emph{pedagogical fitness} for that learner rather than absolute quality.
Crucially, unlike tutoring or role-play uses, treating a simulated learner as an \emph{evaluator} is only meaningful if the persona actually changes the resulting scores; our persona-sensitivity analysis (Section~\ref{sec:persona}) tests exactly this, which prior simulated-learner work does not establish.

\paragraph{Human--AI collaboration and critical use.}
Human--AI teams benefit when their errors differ \citep{steyvers2022,bansal2021complementarity}, but realizing this benefit is hard: people over-rely on automation, and explanations do not reliably fix it \citep{skitka1999,bansal2021explanations,bucinca2021}. 
This raises the question of how much decision authority to leave with a human when an algorithm is ``in the loop'' \citep{green2019}. 
This evidence, however, concerns reliance on AI \emph{assistants} or classifiers---and, more recently, on LLM outputs \citep{si2024large}---rather than on LLMs acting as \emph{evaluators} whose commentary experts weigh while scoring. 
The methodologically closest design is the human--LLM annotation pipeline of \citet{wang2024collab}, where an LLM labels, a verifier scores label quality, and humans re-annotate low-confidence cases. 
We adapt this verify-then-revise structure to expert rating of teaching videos and ask a question this pipeline leaves open: whether the judge's commentary functions as a useful reference signal that experts adopt \emph{critically}, benefiting from it while remaining able to detect when it is wrong. To quantify rater consistency in this workflow, we evaluate reliability using Krippendorff's $\alpha$ \citep{krippendorff2004,hayes2007}.
Table~\ref{tab:related} summarizes the positioning: prior lines each cover some of our axes, but to our knowledge none is at once multimodal, rubric-grounded, and learner-conditioned while also analyzed as a member of a human rating team---and in particular the pairing of  \emph{learner-conditioned evaluation} with \emph{analysis inside a human rating workflow} is previously unoccupied.

\begin{table}[t]
\centering
\small
\caption{Positioning against the closest lines of work. \cmark $=$ a \emph{central}
concern of the cited line; (\cmark) $=$ present but not a central claim; blank $=$ not
addressed. ``Human-team analysis'' means studying the judge's complementarity with, and
critical use by, human experts while they score---not merely agreement with a reference.}
\label{tab:related}
\begin{tabular}{@{}lcccc@{}}
\toprule
 & Multi- & Rubric- & Learner- & Human-team \\
Line of work & modal & grounded & conditioned & analysis \\
\midrule
LLM-judge bias / benchmarks \citep{gu2024survey,lambert2024rewardbench} &  &  &  &  \\
VLM-as-judge \citep{chen2024mllmjudge,xiong2024llavacritic} & \cmark & (\cmark) &  &  \\
Multi-agent judges \citep{du2024debate,chan2024chateval} &  & (\cmark) &  &  \\
Teaching-video generation \citep{chen2025code2video,zhu2025paper2video} & \cmark &  &  &  \\
Educational assessment \citep{wang2023chatgpt,demszky2024feedback} &  & \cmark &  &  \\
Simulated-learner evaluation \citep{lu2024generative} &  & (\cmark) & \cmark &  \\
Simulated learners (role-play) \citep{park2023generative,markel2023gpteach} &  &  & (\cmark) &  \\
\textbf{This work} & \cmark & \cmark & \cmark & \cmark \\
\bottomrule
\end{tabular}
\end{table}

\section{Method}

\subsection{Materials and rubric}
\label{sec:rubric}

We use 12 teaching videos spanning four subject areas (physics, biology, mathematics, and computer science). 
Each subject contains one instructional topic represented by three independently created videos, yielding four topics and twelve videos in total. 
Every video is paired with the course requirement and the target student persona for which it was authored. 
The complete list of videos and topics is provided in Table~\ref{tab:videos} in Appendix~\ref{app:materials}.
The \textbf{course requirement} follows a backward-design specification consisting of a big concept, topic, target Bloom's levels, learning goal, and key learning points. It is provided to the judge together with the video as part of the evaluation input.
The \textbf{student persona} specifies the intended learner through grade level, attention budget, and explicit \texttt{Has}/\texttt{Lacks} prerequisite knowledge. The judge receives the persona together with the video and course requirement. An example is provided in Appendix~\ref{app:materials} (Figure~\ref{fig:example}).
Videos are evaluated using a pedagogy rubric containing 24 metrics grouped into six dimensions (A--F). 
The rubric is derived from learning-science principles, including backward design and Bloom's taxonomy \citep{wiggins2005understanding,bloom1956,anderson2001}, multimedia learning and cognitive load \citep{mayer2014,sweller1988}, cognitive apprenticeship \citep{collins1989cognitive}, and generative and active learning \citep{generativelearning,activelearning}.
Dimension~F evaluates learner adaptability through three metrics: vocabulary appropriateness (F1), prerequisite awareness (F2), and pacing fit (F3). These metrics are scored with respect to the provided student persona.

\paragraph{Metric selection.}

The initial rubric contains 24 metrics derived from learning-science theories. 
We evaluated these metrics on a separate pilot set of high-quality teaching videos, disjoint from the 12 evaluation videos. 
Several metrics exhibited a ceiling effect, with the judge assigning near-uniform top scores across the pilot set, limiting their ability to discriminate between videos.
We therefore retained a curated subset of 10 metrics (Table~\ref{tab:metrics}) for the main human--AI evaluation. 
The selected metrics cover content accuracy and coverage, explanation quality, generative and active learning, and learner adaptability while excluding metrics that saturated on the pilot set. 
Details of the pilot analysis are provided in Appendix~\ref{app:ceiling}.

\begin{table}[t]
\centering
\small
\caption{The 10 evaluated dimensions and their rubric-declared information
modality (\texttt{needs}). Only A1, C4, D2 require visual information.}
\label{tab:metrics}
\begin{tabular}{@{}llll@{}}
\toprule
Code & Dimension & \texttt{needs} & Group \\
\midrule
A1 & Factual accuracy        & transcript + visual & Visual \\
A2 & Topic coverage          & transcript          & Transcript \\
C1 & Examples / analogies    & transcript          & Transcript \\
C4 & Visual explanation power & transcript + visual & Visual \\
D2 & Visual design           & visual              & Visual \\
E1 & Embedded questions      & transcript          & Transcript \\
E2 & Generative prompts      & transcript          & Transcript \\
F1 & Vocabulary appropriateness & transcript       & Transcript \\
F2 & Prerequisite awareness  & transcript          & Transcript \\
F3 & Pacing fit              & transcript + audio  & Transcript \\
\bottomrule
\end{tabular}
\end{table}

\subsection{The three-agent judge}
\label{sec:judge}

\begin{figure}[t]
\centering
\includegraphics[width=0.75\linewidth]{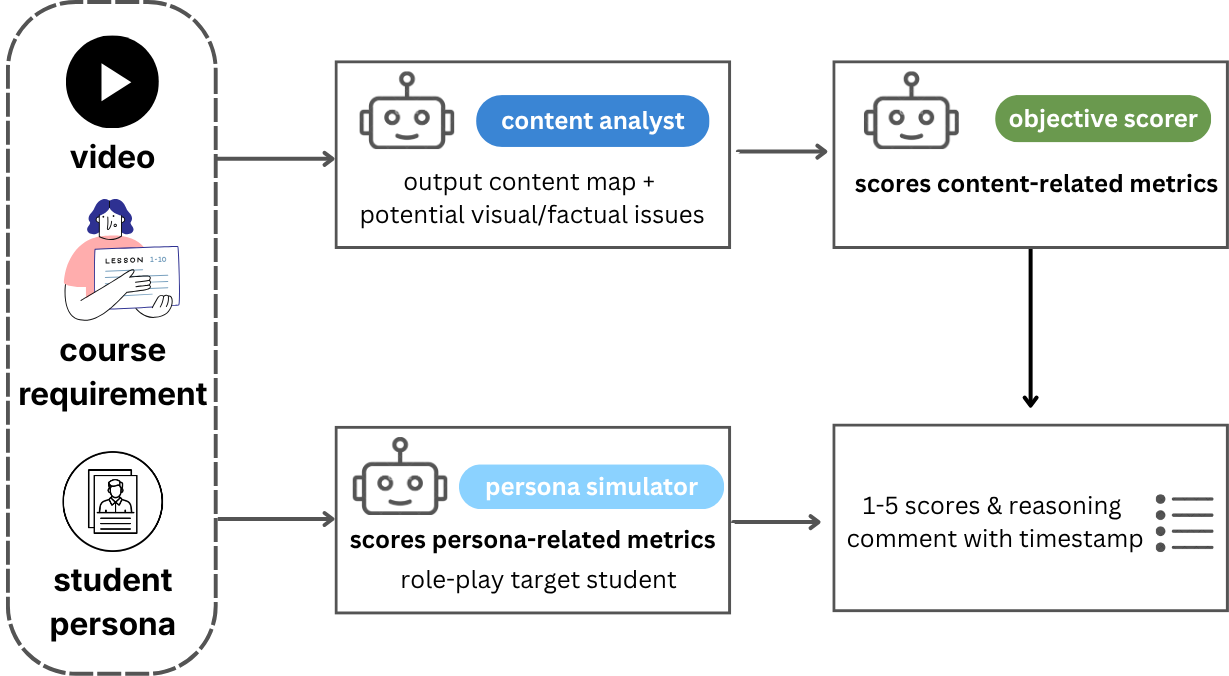}
\caption{Architecture of EduPanel, our three-agent judge. A content-analyst VLM (Agent~1)
watches the video and extracts a content map and potential factual/visual issues;
an objective text scorer (Agent~2) reads Agent~1's report and scores the objective
dimensions (A1, A2); a persona-simulator VLM (Agent~3) role-plays the target
student and scores the experiential dimensions. Dimension routing is set by the
rubric's \texttt{evaluator} field; all agents share one backbone
(\texttt{gemini-3-flash}) and each video is scored over three averaged passes.}
\label{fig:arch}
\end{figure}

\textbf{EduPanel} (Figure~\ref{fig:arch}) takes the video together with its course requirement and student persona, and produces a score from 1 to 5 for each rubric dimension. All three agents use the same backbone (\texttt{gemini-3-flash}), and each video is evaluated over three independent passes whose scores are averaged.
\begin{itemize}
  \item \textbf{Agent 1 --- content analyst} (sees the video): produces a timestamped content map and a list of potential factual/visual issues, including on-screen figure/diagram correctness.
  \item \textbf{Agent 2 --- objective scorer} (text only): scores the objective dimensions from Agent~1's report.
  \item \textbf{Agent 3 --- persona simulator} (video + persona): scores the learner-dependent dimensions using the provided student persona.
\end{itemize}
The assignment of rubric dimensions is specified by the rubric's \texttt{evaluator} field rather than hard-coded into the system: A1 and A2 are assigned to Agent~2, while the remaining dimensions are assigned to Agent~3. 
Both Agent~1 and Agent~3 receive the full video as input, whereas Agent~2 operates only on Agent~1's report.
The contribution of this decomposition is evaluated through the ablation studies in Section~\ref{sec:ablation}.

\subsection{Procedure}
\label{sec:3.3procedure}
\begin{figure}[t]
\centering
\includegraphics[width=0.78\linewidth]{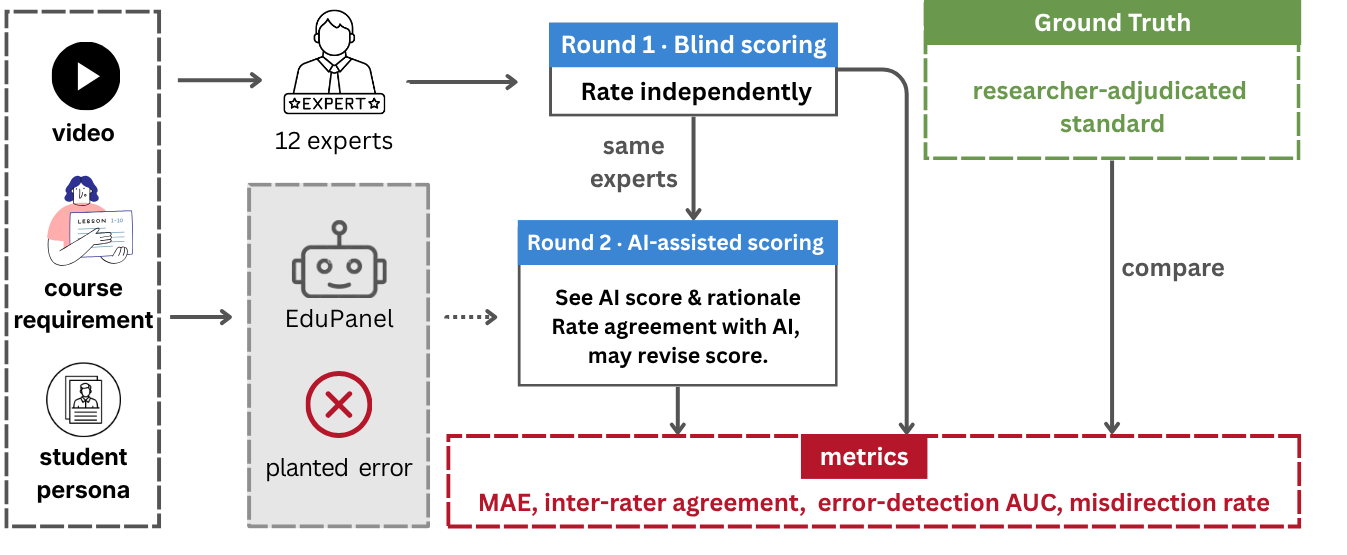}
\caption{Two-round rating procedure. The same 12 experts rate blind, then with AI assistance; some AI scores are deliberately planted errors.}
\label{fig:procedure}
\end{figure}

\paragraph{Participants.} The 12 expert raters were recruited from universities in Taiwan. 
All held at least an undergraduate degree, had prior teaching experience, and were compensated for their time.
\paragraph{Rating protocol.}
Twelve experts rated all 12 videos in two rounds (Figure~\ref{fig:procedure}). 
The two-round verify-then-revise protocol follows the human--LLM collaborative annotation framework of \citet{wang2024collab}, adapted from crowd text annotation to expert evaluation of teaching videos.
In \textbf{Round 1 (blind)} they watched each video and scored the 10 dimensions independently, without seeing the AI. 
In \textbf{Round 2 (AI-assisted)} they saw the judge's score and rationale for each cell, recorded their agreement on a 1--5 scale (\emph{verify}), and could optionally revise their own score. 
Participants were instructed that ``the assistant can be wrong---verify, don't copy''; raters were \emph{not} told that any scores had been deliberately manipulated, nor how many, and planted cells were visually indistinguishable from ordinary AI output.

\paragraph{Ground truth as researcher adjudication.}
\label{sec:gt}

Ground truth (GT) was established by a researcher-adjudicator after reviewing the experts' blind ratings, their discussion, and the AI's commentary.

Because AI feedback was incorporated during adjudication, we additionally evaluate agreement against an AI-free human consensus, computed as the trimmed mean of the 12 blind expert ratings after removing the highest and lowest score for each evaluation cell.

\paragraph{Planted errors.}
To evaluate experts' ability to detect unreliable AI feedback, some AI scores shown in Round~2 were deliberately replaced with planted errors. 
These consisted of both false positives (over-praise) and false negatives (over-criticism), with at most three manipulated cells per video across the ten evaluation dimensions. 
Each planted score differed from the original AI score by at least two points.
Candidate planted errors were screened to ensure that they could be verified from the video itself. 
We excluded cases in which expert consensus was unclear or where identifying the error required off-video domain knowledge. 
This process yielded 26 valid planted errors (9 false positives and 17 false negatives). 
AI outputs without planted errors served as the negative class (the 80 AI-correct cells; see Section~\ref{sec:3.4}) in the detection analysis.
The complete list of planted items and screening decisions is provided in the supplementary material.

\subsection{Measures}
\label{sec:3.4}
Our primary evaluation metrics are mean absolute error (MAE), Pearson correlation ($r$), and Krippendorff's $\alpha$. 
We additionally report the within-1 rate relative to GT as a secondary metric because it is less sensitive to differences in score variance.
For the planted-error analysis, we separately evaluate error detection and behavioral response. 
Error detection is measured by ROC AUC computed from agreement ratings on planted and unmodified AI outputs, while behavioral response is measured by the misdirection rate, namely whether revised ratings move toward the planted score and away from the reference.
We report cluster-bootstrap 95\% confidence intervals by resampling videos ($B=5000$). 
Additional robustness analyses, including leave-one-video-out evaluation and ordinal alternatives to the primary metrics, are provided in Appendix~\ref{app:robust}.
Different analyses use different subsets of the 120 evaluation cells: RQ1 evaluates agreement on all 120 cells; the expert-assistance analysis uses the 86 completely unmodified (non-planted) cells; and the planted-error analysis compares the 26 valid planted errors (excluding 8 of the 34 originally manipulated cells that were screened out as invalid) against the 80 AI-correct cells (defined as non-planted cells where the AI score is within 1 point of the ground truth, leaving 6 cells with larger natural discrepancies excluded).


\subsection{Reproducibility and release}
EduPanel is released as an open-source pipeline together with the full agent prompts, pedagogy rubric, per-dimension routing, per-video course requirements and personas, raw outputs from all three passes, the complete planted-error annotations, and de-identified blind and AI-assisted ratings (Appendix~\ref{app:repro}).
All judge runs use \texttt{temperature}=0, \texttt{seed}=42, and a pinned \texttt{gemini-3-flash-preview} model snapshot.
The released outputs and ground-truth labels support recomputation of all quantitative analyses.
Because the ground truth was produced through human adjudication, its construction is documented in Section~\ref{sec:gt} rather than reproduced by script.
We also release the per-pass outputs so that the reported analyses can be reproduced without re-querying the closed model endpoint.
\section{Results}

\subsection{RQ1 --- EduPanel's evaluation reliability and error profile}
\label{sec:rq1}

\begin{figure}[t]
\centering
\includegraphics[width=0.78\linewidth]{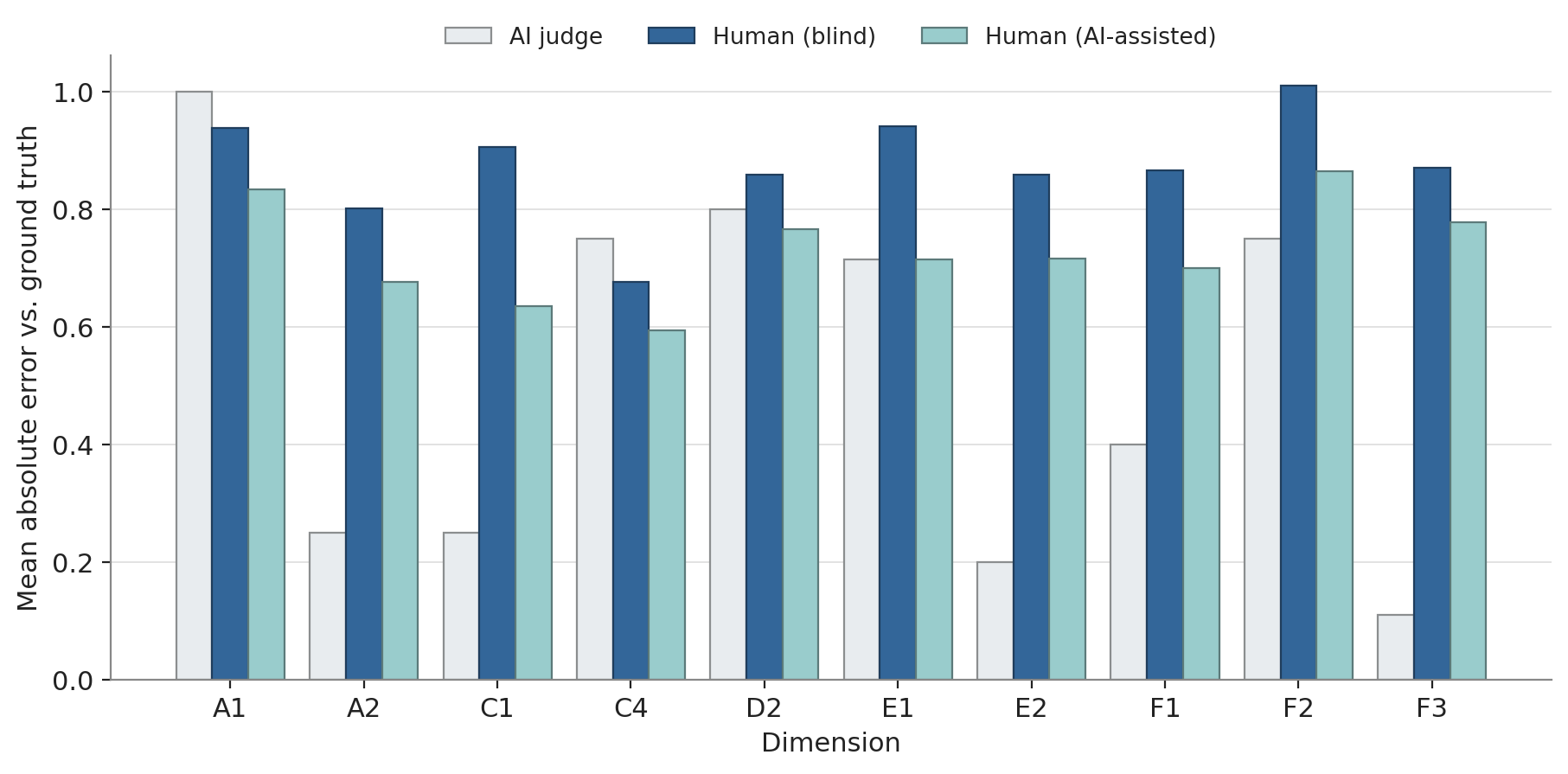}
\caption{Per-dimension MAE against GT on the 86 non-planted cells. EduPanel is most accurate on transcript-grounded dimensions and least accurate on visually grounded ones. AI-assisted human ratings are closer to GT than blind ratings on nearly every dimension. Each dimension contains at most 12 cells (86 in total after excluding planted cells) and should therefore be interpreted descriptively.}
\label{fig:permetric}
\end{figure}

\paragraph{Consistency with human raters.}
EduPanel achieves consistency comparable to a typical human rater.
Against the AI-free trimmed human consensus over all 120 cells, EduPanel achieves an MAE of 0.85, comparable to the median blind expert (MAE = 0.87; range 0.58--1.00).\footnote{Experts are compared against a consensus they helped form, whereas EduPanel is evaluated externally against the same reference.}
Agreement against the adjudicated ground truth is higher (MAE = 0.55; within-1 = 92\%). Because the GT incorporated AI commentary during adjudication (Section~\ref{sec:gt}), we treat agreement against the AI-free human consensus as the primary comparison when evaluating EduPanel against human raters.
Importantly, the adjudicated GT did not simply reproduce the AI's outputs: it differed from the judge on 47\% of evaluation cells, including 8\% that differed by two points or more.
Blind inter-rater agreement is moderate ($\alpha=0.38$; Section~\ref{sec:assist}), so ``human-level'' here refers to agreement with a group of experts whose own ratings are not perfectly consistent.
Per-dimension errors (Figure~\ref{fig:permetric}) reveal a clear modality effect: EduPanel performs best on transcript-grounded dimensions (A2, C1, E2, F3) and least accurately on visually grounded ones (A1, C4, D2). We revisit this pattern in the expert-assistance analysis (Section~\ref{sec:assist}).

\begin{figure}[t]
\centering
\includegraphics[width=0.75\linewidth]{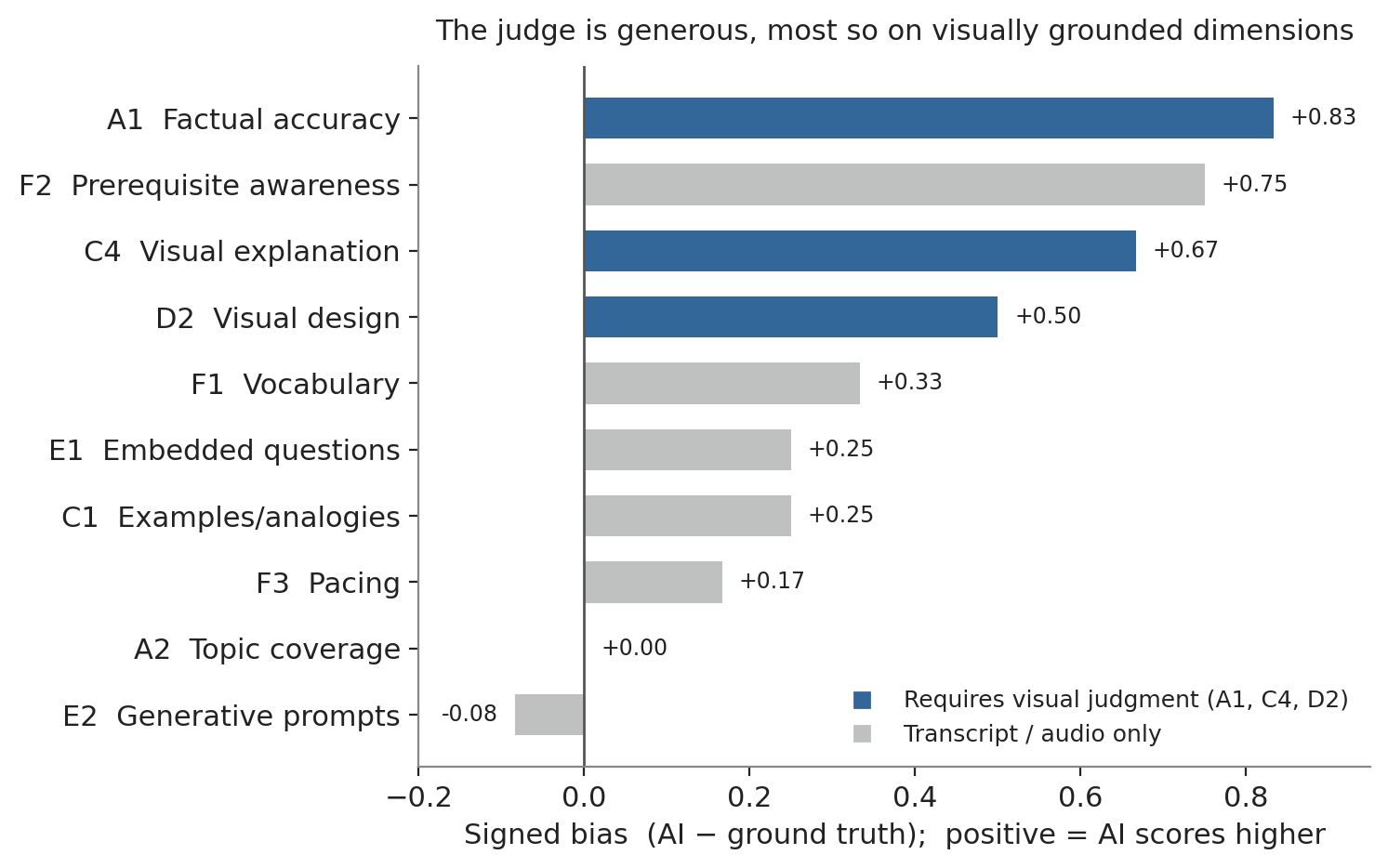}
\caption{
Per-dimension signed bias (AI$-$GT). Visually grounded dimensions (A1, C4, D2) exhibit larger positive bias than transcript-grounded dimensions (group means +0.67 vs. +0.24). Dashed lines indicate group means. Each dimension contains 12 evaluation cells and should therefore be interpreted descriptively.
}
\label{fig:bias}
\end{figure}

\paragraph{A modality-graded leniency on the deployed backbone LLM.}
On the deployed Gemini backbone, the judge exhibits a systematic positive bias that varies across information modalities (Figure~\ref{fig:bias}).
The three visually grounded dimensions show an average signed bias of +0.67, compared with +0.24 for transcript-only dimensions.
The gradient tracks \emph{modality}, not which agent scores the dimension (the objective and subjective agent groups differ by only $+0.06$), and one ablation ties it to the video itself: with no picture to judge (transcript-only, Section~\ref{sec:ablation}) the visual leniency vanishes and reverses ($+0.67\!\rightarrow\!-0.45$). 
This modality-dependent pattern is not observed on the GPT backbone (Table~\ref{tab:backbone}), suggesting that it may reflect characteristics of the deployed Gemini model rather than a universal property of LLM judges.

\paragraph{Complementarity.}
Aggregate agreement does not capture whether the judge identifies different errors from human experts.
EduPanel occasionally identified valid problems that no expert initially noticed. 
For example, in a meiosis video it correctly flagged that the narration places DNA replication in Prophase~I, whereas DNA replication occurs during the S~phase of Interphase. All 12 blind experts accepted the explanation as essentially correct, and one explicitly commented that they lacked the domain knowledge to judge the issue.
This pattern was not unique to a single example. Across the non-planted cells, we identified 16 cases in which GT differed from the blind-expert mean by at least 0.8 points toward EduPanel's evidenced judgment, suggesting that the judge sometimes identifies issues overlooked by the expert group.
Because some of these cases may reflect disagreement among experts rather than uniquely AI-discovered errors, we treat 16 as an upper-bound estimate of complementary cases. The complete list, together with the experts' comments, is provided in the supplementary material.

\paragraph{Cross-backbone robustness.}
\label{sec:xbackbone}
Re-scoring the 12 teaching videos with a GPT backbone (using the same pipeline with sampled video frames) yields two findings (Table~\ref{tab:backbone}).
First, overall evaluation accuracy is robust across backbones. Against the AI-free human consensus, Gemini and GPT achieve similar performance (MAE = 0.85 vs.\ 0.87).
Second, the modality-dependent signed bias is backbone-dependent. Gemini exhibits systematic positive bias (visual: +0.67; transcript: +0.24), whereas GPT shows no net visual bias (+0.06), although this group mean averages over dimensions of opposite sign (A1 +0.66).
Only the positive bias on the visual factual-accuracy dimension (A1) appears consistently across both backbones (+0.83 vs.\ +0.66).
Taken together, these results indicate that overall evaluation accuracy transfers across backbones, whereas the modality-dependent bias does not.

\begin{table}[t]
\centering
\small
\caption{Cross-backbone comparison using the same three-agent pipeline on the 12 teaching videos (GPT uses sampled video frames). Overall evaluation accuracy is similar across backbones when measured against the AI-free human consensus, whereas the modality-dependent signed bias is observed only on the deployed Gemini backbone.}
\label{tab:backbone}
\begin{tabular}{@{}lcccc@{}}
\toprule
Backbone & MAE vs.\ GT & MAE vs.\ consensus & Bias visual & Bias transcript \\
\midrule
\texttt{gemini-3-flash} (deployed) & $0.55$ & $0.85$ & $+0.67$ & $+0.24$ \\
GPT (cross-backbone)               & $1.00$ & $0.87$ & $+0.06$ & $+0.02$ \\
\bottomrule
\end{tabular}
\end{table}

\subsection{RQ2 --- Validating EduPanel's design: architecture ablation}
\label{sec:ablation}
We evaluate the contribution of each design component through four ablations that modify one component at a time while keeping the remainder of the pipeline unchanged (Table~\ref{tab:ablation}).
The four configurations respectively (1) collapse the three agents into a single model call (\emph{monolithic}), (2) remove the content-analysis agent, (3) replace the learner persona with a neutral persona, and (4) remove video input (\emph{transcript-only}).
Performance changes ($\Delta$MAE) are reported relative to the deployed full pipeline (MAE = 0.55).

\paragraph{Video is indispensable.}
Removing the video causes the largest performance degradation among all ablations. Transcript-only MAE rises $0.55\!\rightarrow\!1.07$ (within-1 $92\%\!\rightarrow\!72\%$) and its scores flatten toward the scale center (SD $0.66$, $r$ vs.\ GT $0.23$), with the visual dimensions defaulting to a neutral score. 
The visual leniency of Section~\ref{sec:rq1} even flips sign ($+0.67\!\rightarrow\!-0.45$), confirming that the generosity is a property of actually watching the video, not a text artifact. 
Removing visual input substantially degrades evaluation performance, indicating that transcript-only judging cannot recover the information provided by the video.

\paragraph{Persona conditioning selectively affects learner-adaptability dimensions.}
Replacing the target learner with a neutral persona selectively degrades the learner-adaptability dimensions (F-family MAE: 0.42$\rightarrow$0.90), with all three F dimensions worsening (F2: 0.75$\rightarrow$1.17). In contrast, the objective content dimensions remain largely unchanged (A-family MAE: 0.67$\rightarrow$0.72), with differences comparable to run-to-run variation.
A within-batch re-run of the full pipeline gave MAE = 0.60, so we treat differences within ±0.05 MAE as run-to-run variation (per-pass stability is reported in Table~\ref{tab:retest}).

\paragraph{The monolithic judge compresses score variance.}
Collapsing the three agents into a single model call leaves mean error essentially unchanged (MAE $0.55$ vs.\ $0.55$; within-1 $95\%$ vs.\ $92\%$). 
However, the monolithic judge uses a substantially narrower portion of the rating scale (SD $0.77$ vs.\ GT's $1.22$), never assigning the highest score, whereas the full pipeline produces a score distribution closer to the reference (SD $1.44$). 
The full pipeline also achieves a higher correlation with GT ($r=0.84$ vs.\ $0.78$).
Among these differences, score variance is the most robust: the SD gap remains significant under video-level bootstrap (95\% CI $[+0.52,+0.76]$), whereas the correlation difference is suggestive but not conclusive (95\% CI $[-0.01,+0.17]$).

\paragraph{Effects of agent decomposition.}
Removing the content-analysis agent has little effect on aggregate accuracy (+0.06, comparable to run-to-run variation) but eliminates the intermediate evidence produced by the full pipeline, including timestamped content maps, coverage checklists, and identified factual and visual issues.
The monolithic configuration also substantially reduces the signed bias ($+0.37\rightarrow+0.04$) while increasing the error on the learner-adaptability dimension F3 (MAE $0.17\rightarrow0.86$).

\begin{table}[t]
\centering
\small
\caption{
Architecture ablation on the 12 teaching videos (120 evaluation cells). All configurations use the same \texttt{gemini-3-flash} backbone and evaluation protocol, differing only in the removed component. $\Delta$MAE is measured relative to the deployed full pipeline (MAE = 0.55). A-family denotes objective content dimensions (A1--A2), and F-family denotes learner-adaptability dimensions (F1--F3). Score SD measures the spread of each configuration's predictions.
}
\label{tab:ablation}
\begin{tabular}{@{}lccccccc@{}}
\toprule
Configuration & MAE $\downarrow$ & $\Delta$MAE & within-1 $\uparrow$ & A-fam & F-fam & $r$ vs.\ GT $\uparrow$ & score SD \\
\midrule
Full 3-agent (deployed) & $0.55$ & --- & $92\%$ & $0.67$ & $0.42$ & $\mathbf{0.84}$ & $1.44$ \\
Monolithic              & $0.55$ & $0.00$ & $95\%$ & $0.61$ & $0.61$ & $0.78$ & $0.77$ \\
No content-analyst      & $0.61$ & $+0.06$ & $93\%$ & $0.72$ & $0.56$ & $0.81$ & $1.32$ \\
Persona-ablated         & $0.75$ & $+0.20$ & $87\%$ & $0.72$ & $0.90$ & $0.74$ & $1.34$ \\
Transcript-only         & $1.07$ & $+0.52$ & $72\%$ & $0.94$ & $1.19$ & $0.23$ & $0.66$ \\
\bottomrule
\end{tabular}
\end{table}

\subsection{RQ2 --- Validating learner-conditioned evaluation: persona sensitivity}
\label{sec:persona}

If learner conditioning is functioning as intended, changing the target learner should change the evaluation. To test this, we conducted a persona-sensitivity study on a separate set of 32 teaching videos. Both EduPanel and a non-expert participant pool scored each video twice, once assuming a middle-school learner and once assuming a first-year university learner. We then compared the magnitude and direction of the resulting score changes.

\textbf{EduPanel responds to changes in the target learner.}
Changing the target persona from a middle-school learner to a first-year university learner increases the fitness score by $\Delta=+1.43$ for vocabulary and $\Delta=+1.47$ for prerequisite knowledge (Table~\ref{tab:persona}). The score shifts point toward better fit on 25/32 and 23/32 videos, respectively. Human raters show the same directional pattern, but with substantially smaller shifts ($+0.57$ for both dimensions).

The judge's score changes also track human judgments across videos. The pooled video-level correlation between judge and human score changes is $0.37$, with the strongest agreement on vocabulary ($r=0.41$), followed by prerequisite knowledge ($r=0.20$) and pacing ($r=0.19$).

The effect is not uniform across all learner-adaptability dimensions. 
Vocabulary and prerequisite awareness respond consistently to persona changes, whereas pacing changes only modestly ($\Delta=+0.36$) and shows the expected direction on 15 of 32 videos. 
Accordingly, we make no claim that pacing is reliably learner-conditioned.

This finding is consistent with the architecture ablation (Section~\ref{sec:ablation}), where removing the persona selectively increases error on the learner-adaptability dimensions (F-family MAE $0.42\rightarrow0.90$) while leaving the objective dimensions largely unchanged.

\begin{table}[t]
\centering
\small
\caption{
Persona sensitivity on the learner-adaptability dimensions (F1--F3). Both EduPanel and the human participant pool rated the same 32 videos under two target personas (middle-school and first-year university learner). $\Delta$ denotes the mean score change (university $-$ middle-school), where positive values indicate that the video is judged to better fit the more advanced learner. ``Correct-dir.'' reports the number of videos for which EduPanel shifts in the expected direction.
}
\label{tab:persona}
\begin{tabular}{@{}lccc@{}}
\toprule
Dimension & Judge $\Delta$ & Human $\Delta$ & Judge correct-dir. \\
\midrule
F1 jargon        & $+1.43$ & $+0.57$ & $25/32$ ($78\%$)\\
F2 prerequisite  & $+1.47$ & $+0.57$ & $23/32$ ($72\%$)\\
F3 pacing        & $+0.36$ & $+0.20$ & $15/32$ ($47\%$) \\
\bottomrule
\end{tabular}
\end{table}

\subsection{RQ3 --- EduPanel in an expert workflow: does it help experts?}
\label{sec:assist}
EduPanel provides measurable benefits when used as an expert-assistance tool.
On the 86 non-planted cells, where the judge presents its original outputs, expert accuracy improved after viewing the judge's scores and explanations. Mean MAE decreased from 0.87 (blind) to 0.73 (AI-assisted), an improvement of 0.14 (95\% cluster-bootstrap CI $[0.10,0.19]$), with 8 of the 12 experts improving and none worsening beyond run-to-run variation.
Expert agreement also increased. Krippendorff's $\alpha$ rose from 0.38 to 0.50 ($\Delta=+0.12$, 95\% CI $[0.06,0.17]$), while the mean per-cell standard deviation decreased from 0.98 to 0.85.
Both improvements remain directionally consistent under all five ground-truth construction rules (Table~\ref{tab:gtsens}). 
However, the magnitude of the MAE improvement decreases when evaluated against AI-free reference rules, from +0.14 under the adjudicated GT to +0.08--+0.10 under AI-free consensus references (and +0.05 under the mode rule).
Because the study follows a within-subject blind-then-assisted design without a no-AI re-rating control, the observed improvement should be interpreted as associational rather than causal. Nevertheless, the increase in inter-rater agreement does not depend on the GT construction and therefore provides independent evidence that EduPanel helps experts reach more consistent judgments.

\begin{table}[t]
\centering
\small
\caption{
Sensitivity of the expert-assistance results to five ground-truth construction rules. Blind and assisted MAE are reported under each rule, together with the resulting improvement ($\Delta$MAE = blind $-$ assisted). Detection performance is summarized by ROC AUC and misdirection rate.
}
\label{tab:gtsens}
\begin{tabular}{@{}lrrrrr@{}}
\toprule
GT rule & Blind & Assisted & $\Delta$MAE & AUC & Misdir. \\
\midrule
Curated (adjudicated) & 0.87 & 0.73 & $+0.14$ & 0.77 & 17\% \\
Mean (rounded)        & 0.79 & 0.69 & $+0.10$ & 0.77 & 14\% \\
Median                & 0.76 & 0.68 & $+0.08$ & 0.77 & 15\% \\
Mode                  & 0.78 & 0.73 & $+0.05$ & 0.78 & 17\% \\
Trimmed mean          & 0.78 & 0.68 & $+0.10$ & 0.77 & 15\% \\
\bottomrule
\end{tabular}
\end{table}

\subsection{RQ3 --- Verifying EduPanel's errors: operational safety}
\label{sec:safety}
Beyond improving expert ratings, a useful judge should also allow experts to recognize when it is wrong. We therefore evaluate both error detection and subsequent revision behavior.

\begin{figure}[t]
\centering
\includegraphics[width=0.78\linewidth]{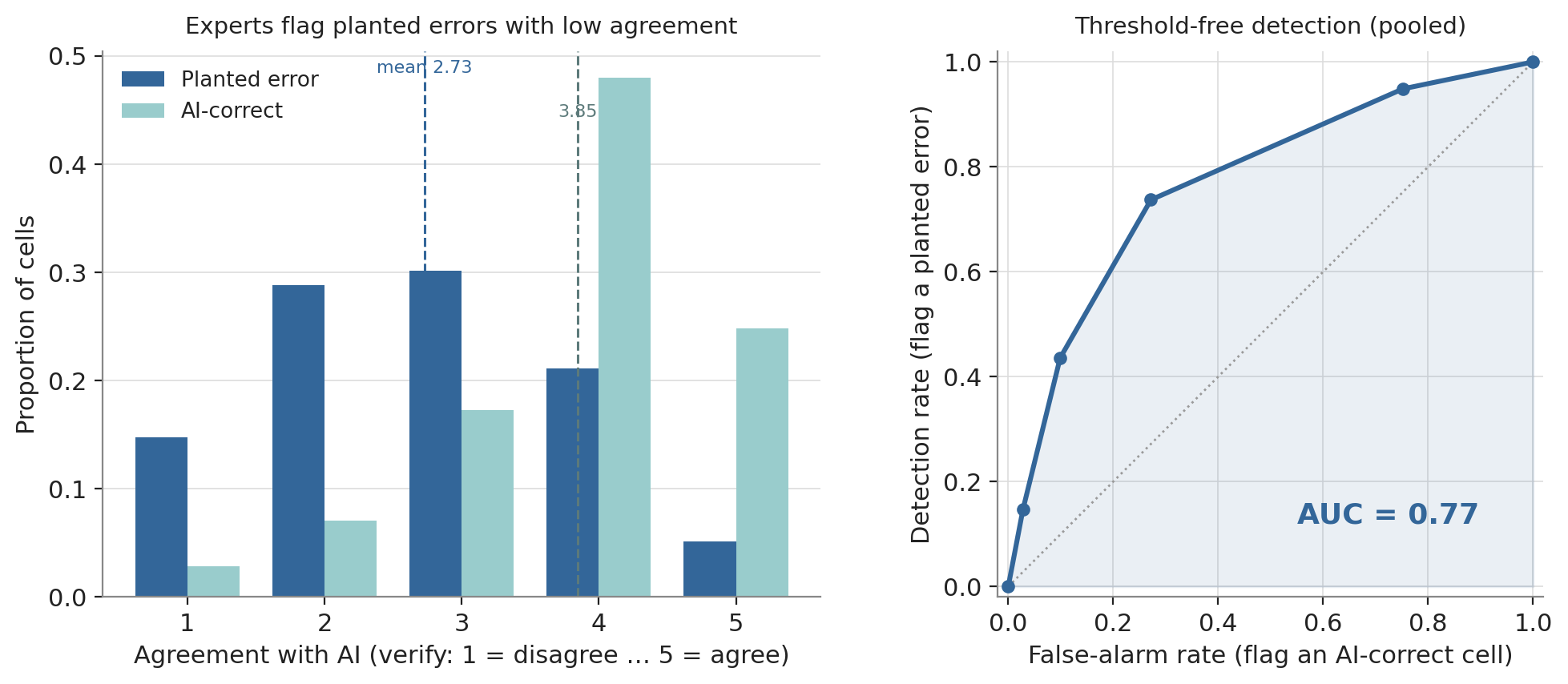}
\caption{Verify-agreement distributions for planted errors and AI-correct outputs. Experts are expected to disagree with planted errors and agree with AI-correct outputs. Their separation is summarized by the ROC AUC.}
\label{fig:sdt}
\end{figure}

\textbf{Error detection.}
Experts assigned substantially lower agreement to planted errors than to AI-correct outputs (mean verify score 2.73 vs.\ 3.85; Figure~\ref{fig:sdt}). This yields a pooled ROC AUC of 0.77 (95\% cluster-bootstrap CI $[0.74,0.80]$), meaning that a planted error receives lower agreement than a correct AI output 77\% of the time.
Detection performance is robust across all five GT construction rules (Table~\ref{tab:gtsens}). Overall catch rates (verify $\le2$) are 51\% for false positives, 40\% for false negatives, and 44\% overall, with a 10\% false-alarm rate on AI-correct outputs. Detection also increases with planted-error magnitude (29\%, 41\%, and 55\% for gaps of 2, 3, and 4 points, respectively).

\textbf{Detection does not imply revision.}
Experts rarely revised their ratings. The misdirection rate was 17\% (of revised ratings on planted-error cells), compared with a constructive-adoption rate of 16\% (of all individual rating instances on AI-correct outputs). 
Overall rating-level revision rates remained similarly low for correct and incorrect AI outputs: only 17\% of the 960 individual rating instances on AI-correct cells were revised, compared with 21\% of the 312 instances on incorrect (planted) cells. 
At the cell level, these revisions were highly clustered, occurring on only 19\% of the 80 AI-correct cells (i.e., 81\% of these cells had zero expert revisions). 
Together with the high detection AUC, these findings suggest that experts often recognize problematic AI outputs without necessarily changing their own ratings.


\section{Discussion}

\paragraph{Beyond multimodal judging.}
Our contribution is not simply a new judge, but a different evaluation objective.
Rather than estimating a single notion of instructional quality, EduPanel evaluates pedagogical fitness for a specified learner.
The conditioning is substantive: changing the learner changes the evaluation, while removing the learner profile selectively degrades learner-adaptability dimensions.

\paragraph{Experimental insights.}
Three findings emerge.

First, evaluation quality depends on information modality.
EduPanel performs well on transcript-grounded dimensions but is more lenient on visually grounded ones.
Because this pattern is absent on the GPT backbone, it appears specific to Gemini rather than a general property of LLM judges.

Second, the three-agent architecture improves calibration more than aggregate accuracy.
A monolithic judge achieves comparable MAE but compresses the score distribution, whereas the full pipeline preserves greater score dispersion and provides interpretable intermediate evidence.

Third, persona conditioning selectively affects learner-adaptability dimensions, indicating that learner representations actively guide evaluation.

\paragraph{Implications for deployment.}
Our workflow study suggests that EduPanel should assist experts rather than replace them.
Experts improved both accuracy and consistency after seeing the judge, while still identifying many incorrect AI outputs.
Notably, this discernment occurred mainly during verification rather than score revision: experts often recognized problematic outputs but rarely changed their ratings.
Thus, disagreement can serve as an informative signal for additional review, especially on visually grounded dimensions where the judge is least reliable.

As AI-generated educational videos scale, scalable evaluation becomes equally important.
Our results suggest that learner-conditioned multimodal judges can support experts, but should currently serve as decision-support systems rather than autonomous evaluators.


\section{Limitations}
\label{sec:limitations}

Our findings should be interpreted within the scope of the present study.

\emph{Ground truth construction.}
The reference standard is based on adjudication by a single researcher who incorporates AI commentary. Consequently, agreement with the adjudicated GT does not constitute an independent validation of the judge itself.
To reduce this dependence, we additionally report results against an AI-free human consensus and evaluate robustness under five alternative GT construction rules.

\emph{Causal interpretation of expert assistance.}
The expert study adopts a within-subject blind-then-assisted design without a no-AI re-rating control.
Consequently, the observed improvement in expert accuracy cannot be cleanly separated from practice effects, scale familiarization, or repeated exposure.
Establishing a causal effect of AI assistance therefore requires an additional controlled condition.

\emph{Scale and statistical power.}
The evaluation benchmark contains 12 videos and 10 analyzed rubric dimensions.
Accordingly, per-dimension results should be interpreted as descriptive trends rather than precise estimates, and our conclusions are limited to this curated benchmark rather than teaching-video evaluation in general.

\emph{Scope of learner conditioning.}
Persona sensitivity is evaluated using only two learner profiles and a crowd reference instead of expert raters.
Furthermore, while the judge internally adjusts scores in the expected direction for both vocabulary and prerequisite dimensions, its co-variation with human-assessed shifts is moderate only for vocabulary adaptation, remaining weak for prerequisite awareness and especially pacing.
Whether learner conditioning generalizes to a broader range of learner characteristics remains an open question.

\emph{Scope of the architectural findings.}
The proposed three-agent architecture improves transparency, controllability, and persona-dependent evaluation, but provides only limited gains in aggregate accuracy over a monolithic judge on this benchmark.
Likewise, the observed modality-dependent leniency appears specific to the deployed Gemini backbone and should not be interpreted as a general property of LLM judges.

Overall, these limitations delineate the scope of our conclusions and motivate future work on larger-scale benchmarks, stronger ground-truth protocols, controlled studies of AI-assisted evaluation, and broader learner modeling.

\section{Future Work}
\label{sec:future}
Several directions naturally follow from this work.
First, future studies should establish the causal effect of AI assistance through controlled experimental designs and stronger ground-truth protocols, including AI-blind adjudication.
Second, evaluating EduPanel on larger and more diverse benchmarks, additional LLM backbones, and a broader range of learner personas will clarify the generality of the findings.
Finally, extending learner-conditioned evaluation beyond instructional videos to other forms of educational content may help build scalable evaluation pipelines for AI-generated educational materials.

\section{Conclusion}
We presented EduPanel, a multimodal, rubric-grounded judge that evaluates teaching videos relative to their intended learner rather than against a single universal notion of quality.
Across a curated benchmark, EduPanel achieves human-level consistency on some evaluation dimensions, while revealing clear limitations on others, particularly those requiring visual reasoning.
Our workflow study further suggests that learner-conditioned judges can serve as effective decision-support tools: experts improve both their accuracy and agreement while remaining able to recognize many incorrect AI judgments.
As educational content becomes easier to generate, evaluating whether it teaches the right learner may become as important as generating the content itself.

\section*{Acknowledgments}

We thank the twelve expert raters for their time, thoughtful evaluations, and
constructive discussions throughout the study. We are especially grateful to
Chun-Wei Chen for designing the learner personas and for many valuable
discussions on the evaluation protocol. We also thank Yun-Man Hsu for
coordinating and supporting the human-subject study.

\bibliographystyle{plainnat}
\bibliography{references}

\clearpage
\newpage
\appendix
\section{Materials and rubric details}
\label{app:materials}
\begin{table}[t]
\centering
\small
\caption{Teaching videos used in the evaluation. Each instructional topic is represented by three independently created videos.}
\label{tab:videos}
\begin{tabular}{ll}
\toprule
Subject & Topic  \\
\midrule
Physics &
Circular motion \& Kepler's third law \\
Biology &
Mitosis \& meiosis \\
Mathematics &
Triangle centers \\
Computer science &
Overfitting vs. underfitting \\
\bottomrule
\end{tabular}
\end{table}

\paragraph{Ceiling-effect pilot.}
\label{app:ceiling}
The dimension-selection pilot (Section~\ref{sec:rubric}) scored 10 strong human teaching
videos---disjoint from the 12 evaluation videos---with the judge over the full 24-metric
rubric (Table~\ref{tab:ceiling}). Several content-reasoning metrics saturate (A3 logical
structure $4.90\pm0.30$, A6 Bloom alignment $4.80\pm0.40$, C2 scaffolding $4.80\pm0.40$,
C3 concept/procedure balance $4.60\pm0.49$): strong videos are indistinguishable on them,
so they cannot support human--AI agreement analysis. This motivates the curated 10-metric
subset ($\star$) used in the alignment study.

\begin{table}[t]
\centering
\small
\caption{Ceiling-effect pilot on 10 strong human teaching videos: per-metric
mean$\pm$SD over the full 24-metric rubric, with band and whether the metric is
included in the alignment study ($\star$). Bands are partitioned by score mean $\mu$ and SD $\sigma$: CEIL ($\mu\!\geq\!4.6, \sigma\!<\!0.55$), near 
($4.5\!\leq\!\mu\!<\!4.6$ or $\mu\!\geq\!4.6$ with $\sigma\!\geq\!0.55$), mid ($3.7\!\leq\!\mu\!\leq\!4.4$), and frontier ($\mu\!\leq\!3.6$).}
\label{tab:ceiling}
\begin{tabular}{@{}llll@{\hspace{2em}}llll@{}}
\toprule
Code & $\mu\pm\sigma$ & Band & Study & Code & $\mu\pm\sigma$ & Band & Study \\
\midrule
A1 & $4.40\pm1.28$ & mid      & $\star$ & C4 & $3.80\pm0.87$ & mid      & $\star$ \\
A2 & $4.50\pm0.50$ & near     & $\star$ & D1 & $3.80\pm0.75$ & mid      &         \\
A3 & $4.90\pm0.30$ & CEIL     &         & D2 & $3.80\pm0.75$ & mid      & $\star$ \\
A4 & $3.30\pm0.64$ & frontier &         & D3 & $3.60\pm1.11$ & frontier &         \\
A5 & $4.70\pm0.64$ & near     &         & D4 & $4.40\pm0.92$ & mid      &         \\
A6 & $4.80\pm0.40$ & CEIL     &         & E1 & $3.30\pm1.42$ & frontier & $\star$ \\
B1 & $3.80\pm0.60$ & mid      &         & E2 & $3.00\pm1.10$ & frontier & $\star$ \\
B2 & $4.00\pm0.89$ & mid      &         & E3 & $3.60\pm0.66$ & frontier &         \\
B3 & $4.30\pm0.90$ & mid      &         & F1 & $4.10\pm0.83$ & mid      & $\star$ \\
B4 & $4.30\pm0.90$ & mid      &         & F2 & $4.20\pm0.87$ & mid      & $\star$ \\
C1 & $4.10\pm0.94$ & mid      & $\star$ & F3 & $3.80\pm1.08$ & mid      & $\star$ \\
C2 & $4.80\pm0.40$ & CEIL     &         &    &               &          &         \\
C3 & $4.60\pm0.49$ & CEIL     &         &    &               &          &         \\
\bottomrule
\end{tabular}
\end{table}

\paragraph{Full rubric and dimension routing.}
The full rubric contains 24 metrics in six dimensions (A--F), each with 1--5 anchors and
an \texttt{evaluator} field naming the scoring agent; dimension~F is scored relative to
the target persona. Table~\ref{tab:routing} lists the ten alignment-study dimensions and
their routing. Agent~1 (content analyst) and Agent~3 (persona simulator) are
vision--language models with full video access; Agent~2 (objective scorer) is text-only
and consumes Agent~1's report.

\begin{table}[h]
\centering
\small
\caption{Rubric-declared routing for the ten evaluated dimensions
(\texttt{evaluator} field). A1--A2 are scored by the objective agent; the
remaining eight by the persona-simulator agent.}
\label{tab:routing}
\begin{tabular}{@{}llll@{}}
\toprule
Code & Dimension & Scoring agent & Modality \\
\midrule
A1 & Factual accuracy            & Agent 2 (objective)  & transcript + visual \\
A2 & Topic coverage             & Agent 2 (objective)  & transcript \\
C1 & Examples / analogies       & Agent 3 (persona)    & transcript \\
C4 & Visual explanation power   & Agent 3 (persona)    & transcript + visual \\
D2 & Visual design              & Agent 3 (persona)    & visual \\
E1 & Embedded questions         & Agent 3 (persona)    & transcript \\
E2 & Generative prompts         & Agent 3 (persona)    & transcript \\
F1 & Vocabulary appropriateness & Agent 3 (persona)    & transcript \\
F2 & Prerequisite awareness     & Agent 3 (persona)    & transcript \\
F3 & Pacing fit                 & Agent 3 (persona)    & transcript + audio \\
\bottomrule
\end{tabular}
\end{table}

\paragraph{Example course requirement and persona.}
Figure~\ref{fig:example} shows the two conditioning artifacts for the meiosis video
(Section~\ref{sec:rq1}), lightly condensed. The persona's \texttt{Lacks} list
(``chromosome structure vocabulary; mitosis vs.\ meiosis'') directly operationalizes the
pedagogical-fitness dimensions for this video.

\begin{figure}[h]
\centering
\small
\fbox{\begin{minipage}{0.94\linewidth}
\textbf{\footnotesize COURSE REQUIREMENT}\\[2pt]
{\ttfamily\footnotesize
Big Concept: Heredity\\
Topic: Cell Division and Meiosis\\
Bloom's Taxonomy: understand\\
Learning Goal: (1)~cells can divide; (2)~chromosomes change during
division; (3)~chromosomes replicate and are equally distributed;
(4)~division relates to growth and reproduction.\\
Key Learning Points: cell division; chromosome changes during division;
chromosome replication; equal distribution of chromosomes.
}\\[6pt]
\textbf{\footnotesize STUDENT PERSONA}\\[2pt]
{\ttfamily\footnotesize
Grade: 7th grade\\
Focus Time: 15 minutes\\
Background Knowledge --- Has: everyday ideas that living things grow and
are made of cells; rough sense that cells can split.
Lacks: chromosome structure vocabulary; mitosis vs.\ meiosis; technical
labels beyond plain-language cell division.
}
\end{minipage}}
\caption{The course requirement and student persona supplied to the judge (and to
human raters) for one video (lightly condensed). Both are authored per video; the persona's
\texttt{Has}/\texttt{Lacks} inventory is what renders the F dimensions
scorable relative to a specific learner (Section~\ref{sec:persona}).}
\label{fig:example}
\end{figure}

\section{Ground truth and planted errors}
\label{app:gt}

\paragraph{Ground-truth construction rules.}
The five GT rules in the sensitivity analysis (Table~\ref{tab:gtsens}), all over the 12
per-cell blind expert scores:
\begin{itemize}
  \item \textbf{Curated} --- the researcher's adjudication (Section~\ref{sec:gt}), the
  primary GT; consolidates blind scores and discussion and consults AI commentary.
  \item \textbf{Mean / Median / Mode} --- blind-score mean (rounded half-up), median
  (half up), or mode (ties toward the value nearest the median). All AI-free.
  \item \textbf{Trimmed mean} --- drop one highest and one lowest blind score, then mean
  (robust to outlier raters). AI-free.
\end{itemize}

\paragraph{Planted errors.}
The 26 valid planted cells (9 FP / 17 FN, $\leq3$ per video) each differ from the AI's
real evidenced score by $\geq2$ points. The validity screen is described in Section~\ref{sec:3.3procedure}; the full per-item list (video,
dimension, error type, real vs.\ planted score, gap, screen decision) is in the
supplementary material.

\section{Judge configuration and reproducibility}
\label{app:repro}
All three agents use \texttt{gemini-3-flash} (native MP4 ingestion); the GPT variant for the cross-backbone comparison samples $\sim$48 evenly spaced frames. 
Each video is scored over three independent passes and averaged; run-to-run stability is in Table~\ref{tab:retest}. All scores were collected in a fixed window (snapshot \texttt{gemini-3-flash-preview}, temperature 0, decode seed 42), and we release the full
per-pass outputs---together with the rubric, agent prompts, per-video requirements and
personas, anonymized rater data, planted-error file, and analysis scripts---so results
can be re-derived without re-querying the closed endpoint.

\begin{table}[h]
\centering
\small
\caption{Judge run-to-run reliability: mean per-cell standard deviation across the
three scoring passes (1--5 scale), and the dimensions with the largest pass-to-pass
variance. The deployed Gemini judge is highly stable; the GPT backbone is noisier,
with its instability concentrated on the more subjective/visual dimensions (F2,
C4, A1), consistent with Section~\ref{sec:rq1}. This table reports only the GPT
backbone's run-to-run \emph{stability}; its \emph{accuracy} and bias versus Gemini
are in Table~\ref{tab:backbone}.}
\label{tab:retest}
\begin{tabular}{@{}lcl@{}}
\toprule
Backbone & Mean per-cell SD & Highest-variance dimensions \\
\midrule
\texttt{gemini-3-flash} (deployed) & $\approx 0.13$ & F2, F3 ($\approx 0.20$) \\
GPT (cross-backbone)               & $\approx 0.25$ & F2 ($0.39$), C4 ($0.34$), A1 ($0.31$) \\
\bottomrule
\end{tabular}
\end{table}

\section{Robustness re-analyses}
\label{app:robust}
Three zero-cost re-analyses, all recomputed from the released per-cell scores, check
that the headline results do not depend on modeling or resampling choices.

\paragraph{Rank-based and ordinal treatments.} Replacing the interval treatment of the
ordinal scores with rank/ordinal analogues leaves every conclusion intact. The
judge--GT discrimination is $\rho=0.82$ under Spearman versus $r=0.84$ under Pearson;
the assistance-driven reliability gain is blind $0.36\!\rightarrow\!$ assisted $0.47$
under ordinal Krippendorff $\alpha$ versus $0.38\!\rightarrow\!0.50$ under interval
$\alpha$ (both remain below the conventional $0.667$ threshold).

\paragraph{Leave-one-video-out.} Because the study has only 12 video clusters---below
the range where the cluster bootstrap is reliably calibrated---we additionally drop each
video in turn and recompute the three headline numbers. None is propped up by a single
video: the judge's MAE against GT stays in $[0.51,0.60]$ (full sample $0.55$), the
assistance $\Delta$MAE in $[+0.13,+0.16]$ and positive throughout (full $+0.14$), and
the detection AUC in $[0.76,0.78]$ (full $0.77$).

\paragraph{Retained-dimension variance.} The 10 alignment dimensions were selected for
\emph{judge} score variance on the ceiling pilot (Section~\ref{sec:rubric}); all 10 also
carry substantial \emph{human} blind-score variance (per-dimension SD $1.07$--$1.44$),
so the retained set is not one on which only the judge discriminates. Because the
experts never scored the 14 excluded dimensions, we cannot recompute the selection
itself from human variance; this check bounds, but does not eliminate, the concern that
selection favors dimensions the judge discriminates on (Section~\ref{sec:rubric}).

\end{document}